\documentclass[aps,prd,tightenlines,preprintnumbers,superscriptaddress,twocolumn,notitlepage,nofootinbib,eqsecnum,floatfix]{revtex4-1}
\usepackage{graphicx}
\usepackage[pdftex]{color}
\usepackage{hyperref}
\hypersetup{
    colorlinks=true,       % false: boxed links; true: colored links
    linkcolor=blue,          % color of internal links
    citecolor=blue,        % color of links to bibliography
    filecolor=blue,      % color of file links
    urlcolor=blue           % color of external links
}
\usepackage{amsmath}
\usepackage[utf8]{inputenc}
\usepackage{cases}

\newcommand{\st}[1]{{}}                                 % old text to be deleted
\definecolor{grey}{rgb}{0.6,0.6,0.6}
\newcommand{\amu}{$a_\mu^{\rm HVP}$}
\newcommand{\Pade}{Pad{\'e}}

\newcommand{\Gfit}{$G_{\rm fit}(t)$}
\newcommand{\DeltaIB}{$\delta a_\mu^{{\rm HVP,} m_u \neq m_d}$}

\newcommand{\Pihat}{\widehat\Pi}

\newcommand{\order}{\ensuremath{\text{O}}} %use consistent notation for "order", whatever we decide
\newcommand{\be}{\begin{equation}}
\newcommand{\ee}{\end{equation}}

\newcommand{\ltapprox}{\lesssim} % use AMSmath
\newcommand{\gtapprox}{\gtrsim}  % use AMSmath\]

\begin{document}
\preprint{FERMILAB-PUB-17-486-T}

\widetext

\title{Strong-isospin-breaking correction to the muon anomalous magnetic moment from lattice QCD at the physical point}

%%Author list
\author{B.~Chakraborty}
\affiliation{Jefferson Lab, 12000 Jefferson Avenue, Newport News, Virginia 23606, USA}

\author{C.~T.~H.~Davies}\email{christine.davies@glasgow.ac.uk}
\affiliation{SUPA, School of Physics and Astronomy, University of Glasgow, Glasgow, G12 8QQ, UK}

\author{C.~DeTar} 
\affiliation{Department of Physics and Astronomy, University of Utah, \\ Salt Lake City, Utah, 84112, USA}

\author{A.~X.~El-Khadra}
\affiliation{Department of Physics, University of Illinois, Urbana, Illinois, 61801, USA}
\affiliation{Fermi National Accelerator Laboratory, Batavia, Illinois, 60510, USA}

\author{E.~G\'amiz}
\affiliation{CAFPE and Departamento de F\'{\i}sica Te\'orica y del Cosmos, Universidad de Granada,18071, Granada, Spain}

\author{Steven~Gottlieb}
\affiliation{Department of Physics, Indiana University, Bloomington, Indiana, 47405, USA}

\author{D.~Hatton} 
\affiliation{SUPA, School of Physics and Astronomy, University of Glasgow, Glasgow, G12 8QQ, UK}

\author{J.~Koponen} 
\affiliation{INFN, Sezione di Roma Tor Vergata, Via della Ricerca Scientifica 1, 00133 Roma RM, Italy}

\author{A.~S.~Kronfeld}
\affiliation{Fermi National Accelerator Laboratory, Batavia, Illinois, 60510, USA}
\affiliation{Institute for Advanced Study, Technische Universit\"at M\"unchen, 85748 Garching, Germany}

\author{J.~Laiho}
\affiliation{Department of Physics, Syracuse University, Syracuse, New York, 13244, USA}

\author{G.~P.~Lepage} 
\affiliation{Laboratory for Elementary-Particle Physics, Cornell University, Ithaca, New York 14853, USA}

\author{Yuzhi~Liu} 
\affiliation{Department of Physics, Indiana University, Bloomington, Indiana, 47405, USA}

\author{P.~B.~Mackenzie}
\affiliation{Fermi National Accelerator Laboratory, Batavia, Illinois, 60510, USA}

\author{C.~McNeile}
\affiliation{Centre for Mathematical Sciences, Plymouth University, PL4 8AA, UK}

\author{E.~T.~Neil}
\affiliation{Department of Physics, University of Colorado, Boulder, Colorado 80309, USA}
\affiliation{RIKEN-BNL Research Center, Brookhaven National Laboratory, \\ Upton, New York 11973, USA}

\author{J.~N.~Simone}
\affiliation{Fermi National Accelerator Laboratory, Batavia, Illinois, 60510, USA}

\author{R.~Sugar}
\affiliation{Department of Physics, University of California, Santa Barbara, California, 93016, USA}

\author{D.~Toussaint}
\affiliation{Department of Physics, University of Arizona, Tucson, Arizona, 85721, USA}

\author{R.~S.~\surname{Van de Water}}\email{ruthv@fnal.gov}
\affiliation{Fermi National Accelerator Laboratory, Batavia, Illinois, 60510, USA}

\author{A.~Vaquero} 
\affiliation{Department of Physics and Astronomy, University of Utah, \\ Salt Lake City, Utah, 84112, USA}

\collaboration{Fermilab Lattice, HPQCD, and MILC Collaborations}
\noaffiliation

\vskip 0.25cm
%%%%%%%

\date{\today}

\begin{abstract}
All lattice-QCD calculations of the hadronic-vacuum-polarization contribution to the muon's anomalous magnetic moment to date have been performed with degenerate up- and down-quark masses.  Here we calculate directly the strong-isospin-breaking correction to \amu\ for the first time with physical values of $m_u$ and $m_d$ and dynamical $u$, $d$, $s$, and $c$ quarks, thereby removing this important source of systematic uncertainty.  We obtain a relative shift to be applied to lattice-QCD results obtained with degenerate light-quark masses of \DeltaIB= +1.5(7)\%, in agreement with estimates from phenomenology.
\end{abstract}

\pacs{}
\maketitle

\section{Introduction}
\label{sec:intro}

The anomalous magnetic moment of the muon, $a_\mu \equiv (g_\mu-2)/2$, provides a stringent test of the Standard Model and a sensitive probe of new particles and forces beyond.  
It has been measured by BNL Experiment E821 to a precision of 0.54 parts-per-million~\cite{Bennett:2006fi}, and the experimental result disagrees with Standard-Model theory expectations by more than three standard deviations~\cite{Blum:2013xva}. 
To investigate this discrepancy, the Muon $g-2$ Experiment at Fermilab aims to reduce the experimental error by a factor of four, with a first result competitive with E821 expected within a year~\cite{Grange:2015fou}. 
To identify definitively whether any deviation observed between theory and experiment is due to new particles or forces, the theory error must be reduced to a commensurate level.   

\begin{figure}[tb]
\centering
\includegraphics[width=0.2\textwidth]{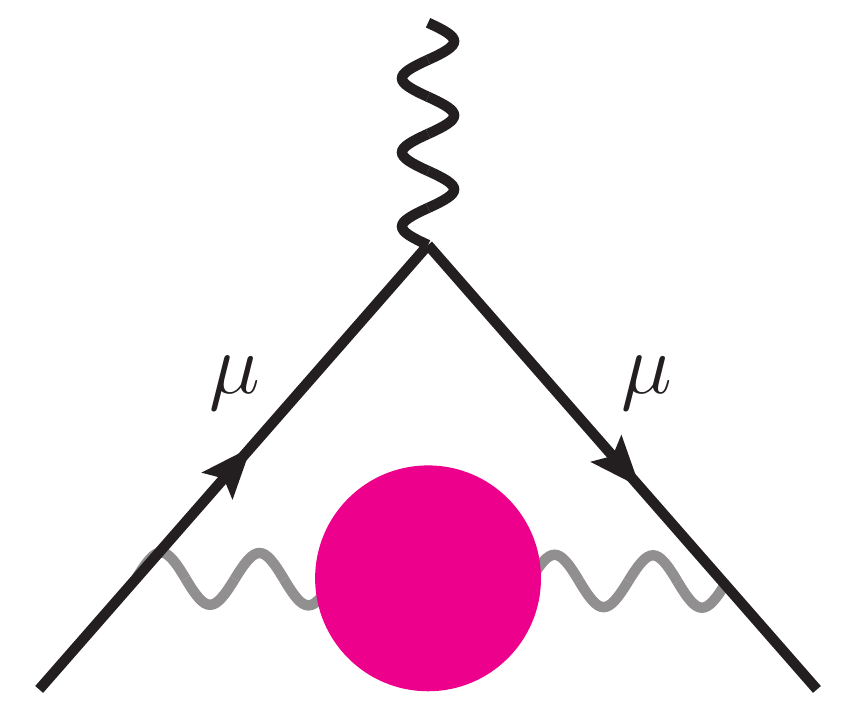}
\caption{
Leading hadronic contribution to $a_\mu$.  The shaded circle denotes all corrections to the internal photon propagator from the vacuum polarization of $u$, $d$, $s$, $c$, and $b$ quarks in the leading one-loop muon vertex diagram.}
\label{fig:HVP}
\end{figure}

The largest source of theory uncertainty is from the \order($\alpha_{\rm EM}^2$) hadronic vacuum-polarization contribution~\cite{Blum:2013xva}, which is shown in Fig.~\ref{fig:HVP} and is denoted by \amu\ throughout this work.
A well-established method for determining this contribution employs dispersion relations combined with experimentally measured electron-positron inclusive scattering cross-section data.  Recent determinations from this approach quote errors of  0.4--0.6\%~\cite{Jegerlehner:2017lbd,KNT17,Davier:2017zfy}.   
Numerical lattice quantum chromodynamics (QCD) provides a complementary, systematic approach for calculating \amu\ directly from first-principles QCD.  Several independent lattice-QCD efforts to obtain \amu\ are ongoing~\cite{Aubin:2006xv,Chakraborty:2016mwy,DellaMorte:2017dyu,BMW17,Lehner:2017kuc}, with errors on recent determinations ranging from about 2--6\%~\cite{Burger:2013jya,Chakraborty:2016mwy,DellaMorte:2017dyu,BMW17}.  The theoretical precision on \amu\ needed to match the target experimental uncertainty is about 0.2\%.

In this Letter, we remove one of the largest systematic errors common to all current lattice-QCD calculations of \amu, namely that arising from the use of degenerate up- and down-quark masses.  We do so by calculating directly the strong-isospin-breaking contribution to the light-quark-connected contribution to \amu. 
Phenomenological estimates suggest that the effect of strong-isospin breaking on \amu\ is about 1\%~\cite{Wolfe:2010gf,Jegerlehner:2011ti,Jegerlehner:2017gek}.
Electromagnetic effects are also not yet included in lattice-QCD calculations of \amu\, and lead to a similar-sized uncertainty~\cite{Hagiwara:2003da,Davier:2009ag}. In order to disentangle quark-mass from electromagnetic effects, we define the strong-isospin-breaking correction using up- and down-quark masses tuned from experimental hadron masses with QED effects removed~\cite{Basak:2016jnn}.

The effect of strong-isospin breaking on the light- and strange-quark connected contributions to \amu\ has been calculated in an exploratory study by the RBC/UKQCD Collaborations~\cite{Boyle:2017gzv} in three-flavor lattice-QCD, with a heavy pion mass of about 340~MeV, and isospin-symmetric sea quarks.  
Preliminary four-flavor results for the strong-isospin-breaking contribution to \amu\ have also been presented by the ETM Collaboration~\cite{Giusti:2017ier} for several pion masses down to $M_\pi \sim 210$~MeV, but with low statistics.
In this Letter, we analyze two QCD gauge-field ensembles recently generated by the MILC Collaboration with four flavors of highly improved staggered (HISQ) sea quarks and very high statistics; see Ref.~\cite{Bazavov:2012xda} for methodology.   
One of the ensembles has fully nondegenerate quark masses with the $u,d,s,$ and $c$ quarks all fixed to their physical values. 
Our calculation is the first determination of \DeltaIB\ at the physical pion mass and with sea isospin breaking.  

This Letter is organized as follows.  We first present our numerical lattice-QCD calculation in Sec.~\ref{sec:LatSims}.  Next, in Sec.~\ref{sec:analysis}, we calculate the strong isospin-breaking correction to \amu\ and discuss the contributions to the systematic error.   We present our final result and compare it with phenomenological estimates and previous lattice-QCD calculations in Sec.~\ref{sec:conclusions}.

\section{Lattice calculation}
\label{sec:LatSims}

We calculate the strong-isospin-breaking correction to \amu\ on two new QCD gauge-field ensembles generated by the MILC Collaboration with four flavors of HISQ quarks~\cite{Follana:2006rc,Bazavov:2012xda}. Table~\ref{tab:ensembles} summarizes key parameters of the configurations.  The two ensembles have the same lattice spacing, which is approximately 0.15~fm, and the same strange- and charm-quark masses, which are both fixed close to their physical values.   With staggered quarks, the pions possess an additional ``taste" quantum number.  Discretization errors from the HISQ action generate $\order(\alpha_s^2 a^2)$ corrections to the squared sea-pion masses of different tastes.  On both ensembles, the mass of the taste-Goldstone $\bar{u}d$ pion is fixed close to Nature's value of $M_{\pi^0} \approx 135$~MeV, which is the mass that the charged pion would have in the absence of electromagnetism.  The root-mean-squared pion mass (averaging over tastes) is about $300$~MeV. 

\begin{table}[tb]
    \caption{Parameters of the QCD gauge-field ensembles.  The first column shows the ratio of the lattice spacing to the gradient-flow scale $w_0$~\cite{Borsanyi:2012zs}.  To convert quantities in lattice-spacing units to GeV, we use $w_0=0.1715(9)$~fm~\cite{Dowdall:2013rya}.  The next columns show the bare lattice up, down, strange, and charm sea-quark masses in lattice-spacing units and the number of configurations analyzed.  The last column gives the taste-Goldstone sea-pion mass in GeV on each ensemble obtained from fits of pseudoscalar-current two-point correlators as in Ref.~\cite{Bazavov:2012xda}. Both ensembles have the same volume $N_s^3 \times N_t = 32^3 \times 48$ in lattice spacing units. \vspace{1mm}}
    \label{tab:ensembles}
\begin{ruledtabular}
\begin{tabular}{lccc} 
$w_0/a$ & $am_u/am_d/am_s/am_c~(\times 10^{2})$ & $N_{\rm conf.}$ & $M_{\pi}$~\!(GeV) \\ 
\hline
1.13215(35) & 0.2426/0.2426/6.73/84.47 & 1902 & 0.1347(7) \\
1.13259(38) & 0.1524/0.3328/6.73/84.47 & 4963 & 0.1346(7)
\end{tabular}
\end{ruledtabular}
\end{table}

The two ensembles differ in one key feature: the values of the light sea-quark masses.  Ensemble 1 is isospin symmetric, with the two light sea-quark masses equal and fixed to $m_l = (m_u + m_d)$/2.  Ensemble 2 features isospin breaking; here the two light-quark masses have the same average light-quark mass as ensemble 1, but the ratio of the light sea-quark masses is fixed to the value of $m_u/m_d$ determined from the MILC Collaboration's study of pion and kaon electromagnetic mass splittings within the quenched approximation of QED~\cite{Basak:2016jnn}.  Because the up and down sea-quark masses on this ensemble each take their physical values, a chiral extrapolation is not needed in our analysis. Comparing results on the two ensembles enables us to quantify the (tiny) effects of sea isospin breaking.  

Our analysis strategy closely follows that of Ref.~\cite{Chakraborty:2016mwy}.
On each ensemble, we calculate vector-current correlators $\langle j_\mu(\mathbf{x},t) j_\mu(\mathbf{0},0) \rangle$ with zero spatial momentum and all four combinations of local and spatially smeared interpolating operators at the source and sink. The smearing function is given in Eq.~(A1) of Ref.~\cite{Chakraborty:2016mwy}, and we employ the same smearing parameters as in that work.
To determine the quark-mass dependence of \amu, we compute correlators with three valence-quark masses $m_q = \{m_u, (m_u+m_d)/2, m_d\}$.  With staggered quarks, the local vector current is not the conserved current, and must be renormalized.  The renormalization factor $Z_{V}^{qq}$ for HISQ quarks, however, has only mild quark-mass dependence so it cancels when the strong-isospin-breaking correction is calculated as a percentage shift.  Additional details on the correlator construction and wavefunction smearings can be found in Ref.~\cite{Chakraborty:2016mwy}. 

We fit the $2\times 2$ matrix of correlators together using the multiexponential parametrization in Eq.~(A6) of Ref.~\cite{Chakraborty:2016mwy}. The fit function includes contributions from both vector and opposite-parity states that arise with staggered valence quarks.  The smeared correlators have smaller overlap with excited states than the local-local correlator, and therefore improve the determination of the energies and amplitudes.     
We fit the correlators over the symmetric time range [$t_{\rm min}, T-t_{\rm min}$], thereby ensuring that the fit describes the correlator over the entire lattice time extent $T$.
To reduce the degrees of freedom in the fit, in practice we average the correlator at times $t$ and $T-t$ and fit only to the lattice midpoint; we also average the smeared-source, local-sink correlator with the local-source, smeared-sink correlator.  
Because our limited number of configurations do not enable us to reliably determine the smallest eigenvalues of the correlation matrix, we employ singular-value-decomposition (SVD) cuts with the values chosen to obtain stable fits with good correlated $\chi^2$ values.  In practice, we replace all eigenvalues below the cut with the value of the SVD cut times the largest eigenvalue; this prescription increases the variance of the eigenmodes associated with the replaced eigenvalues and, thus, the errors on the fit parameters.  We choose the number of states and fit range based on the stability of the ground-state and first-excited-state energies and amplitudes.

For both ensembles and all valence-quark masses, we obtain good correlated fits with stable central values and errors using $t_{\rm min}/a \geq 3$, $N_{\rm states} \geq 3$, and an SVD cut of 0.015, which modifies about 40\% of the eigenvalues of the correlation matrix.  For each of our six fits, the contribution to the $\chi^2$ from the 66 correlator data points ranges from about 45-80.
 Although the lowest-energy states in the vector-current correlators are $I=1$ $\pi\pi$ pairs, we do not see any evidence of such states in our two-point correlator fits.  This is not surprising because there are only a few $\pi\pi$ states below the $\rho$ mass in these correlators, and their amplitudes are suppressed by the reciprocal of the spatial volume.  The ground-state energies for the correlators with $m_q = m_l$ are $E_0 = 776.7(6.5)$~MeV and $E_0 = 779.4(5.1)$~MeV on the $N_f=2+1+1$ and $N_f = 1+1+1+1$ ensembles, respectively; these are statistically consistent with the PDG average for the Breit-Wigner mass $M_{\rho^0} = 775.26(25)$~MeV~\cite{Olive:2016xmw}.

Following Ref.~\cite{Chakraborty:2016mwy}, we reduce the statistical errors in \amu\ by replacing the correlator data at large times by the result of the multiexponential fit.  Although the fit function is appropriate for the periodic lattice temporal boundary conditions, we correct for the finite lattice temporal size by using the infinite-time fit function and doubling the correlator extent to $t=2T$.  We use the fitted correlator above $t^*> 1.5$~fm; with this choice, roughly 80\% of the value of \amu\ comes from the data region.  The values of \amu\ computed with \Gfit\ for $t^*> 1.5$~fm agree within $\sim 1\sigma$ with those computed entirely from data, but with more than ten times smaller statistical errors for $m_q = m_u$.

\section{Analysis}
\label{sec:analysis}

We calculate \amu\ using the method introduced by the HPQCD Collaboration~\cite{Chakraborty:2014mwa}, in which one constructs the $[n,n]$ and $[n,n-1]$ \Pade\ approximants for the renormalized hadronic vacuum polarization function [$\Pihat(q^2)$] from time moments of zero-momentum vector-current correlation functions.   These moments are proportional to the coefficients $\Pi_{j}$ in a Taylor expansion of $\Pihat(q^2)$ around $q^2=0$.  The true result is guaranteed to lie between the $[n,n]$ and $[n,n-1]$ \Pade\ approximants.  We employ the $[3,3]$ \Pade\ approximant for $\Pihat(q^2)$ obtained from the first six Taylor coefficients; the values of \amu\ computed from the $[3,2]$ and $[3,3]$ \Pade\ approximants differ by $0.1 \times 10^{-10}$.

\begin{figure}[tb]
\centering
\includegraphics[width=0.4\textwidth]{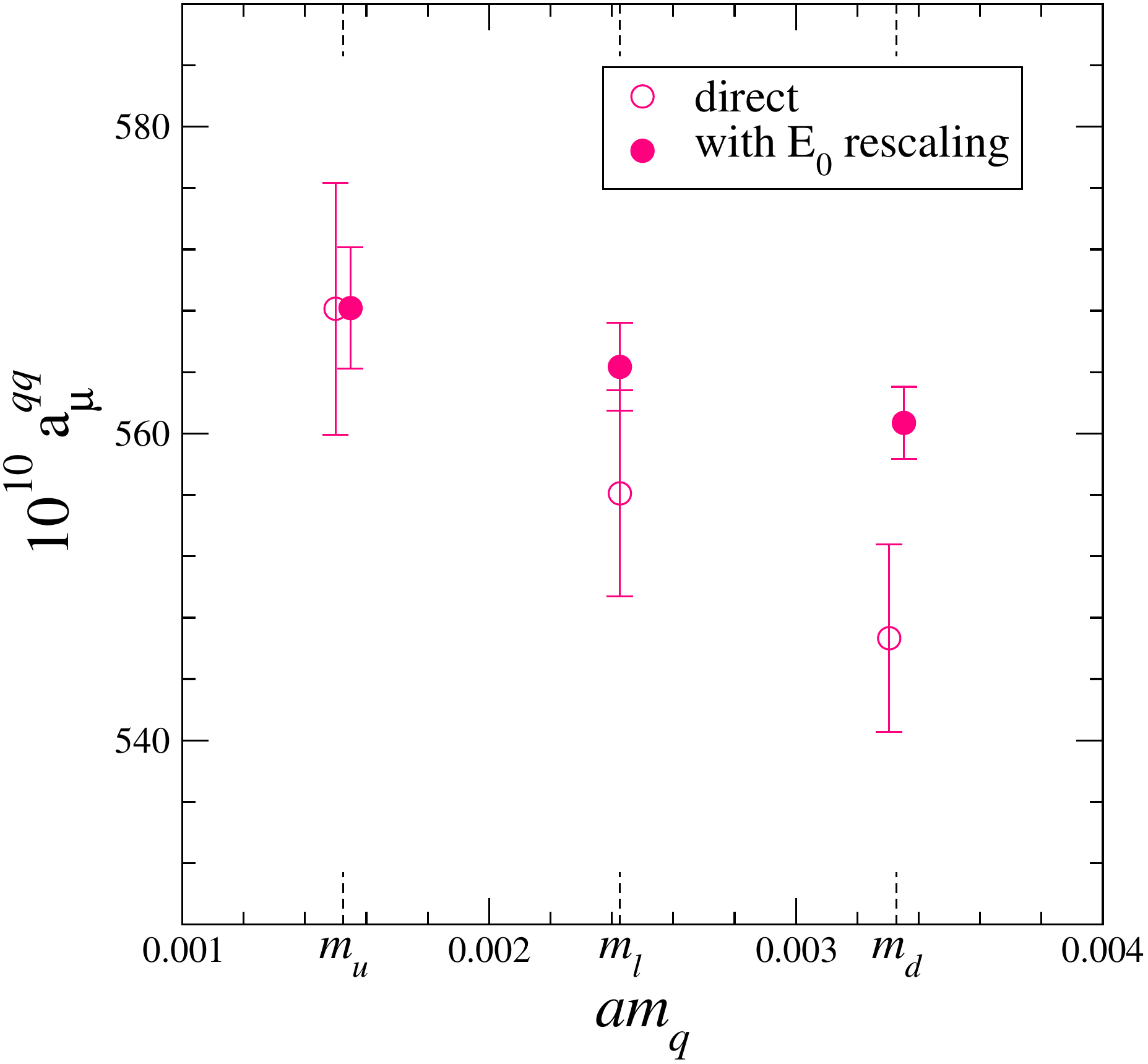} 
\caption{Valence-quark-mass dependence of the light-quark-connected contribution to \amu\ on the $N_f=1+1+1+1$ ensemble without rescaling (open symbols) and with rescaling each $\Pi_j^{(qq)}$ by $(E_0/M_{\rho^0})^{2j}$ (closed symbols).  From left to right, the pairs of data points correspond to $m_u$, $m_l = (m_u+m_d)/2$, and $m_d$; each pair of data points is horizontally offset for clarity.  The values of $a_\mu^{qq}$ include the charge factor $q_u^2 + q_d^2 = 5/9$ appropriate for the isospin-symmetric case.}
\label{fig:amu_vs_mval}
\end{figure}

In Ref.~\cite{Chakraborty:2016mwy}, the $[n,n]$ and $[n,n-1]$ \Pade\ approximants for $\Pihat(q^2)$ are constructed from rescaled Taylor coefficients $\Pi_{j} \times (E_0/M_{\rho^0})^{2j}$, where $E_0$ is the ground-state energy obtained from the two-point correlator fits.  The rescaling was found to reduce the valence-quark-mass dependence of \amu\ because the $\rho$-meson pole dominates the vacuum polarization.  In addition, the rescaling cancels most of the error from the uncertainty on the lattice scale $w_0$, which enters via the muon mass present in the one-loop QED integral for \amu.   Figure~\ref{fig:amu_vs_mval} shows \amu\ on $(1+1+1+1)$-flavor ensemble at the up, down, and average light-quark masses.  The valence-quark-mass dependence is statistically well resolved because the three points are strongly correlated, and is smaller after rescaling.

The physical value of the light-quark-connected contribution to \amu\ is given by the sum of \amu\ with two up quarks in the vector current  and with two down quarks in the vector current weighted by the square of the quarks' electromagnetic charges:
\be
	a_\mu^{\rm phys.} = \frac{4 a_\mu^{uu} + a_\mu^{dd}}{5} \,.
\ee
We then define the absolute shift with respect to the isospin-symmetric value as $\Delta a_\mu^{{\rm HVP,} m_u \neq m_d} \equiv a_\mu^{\rm phys.} - a_\mu^{ll}$, and the relative correction to be
\be
\delta a_\mu^{{\rm HVP,} m_u \neq m_d} = \frac{\Delta a_\mu^{{\rm HVP,} m_u \neq m_d}}{a_\mu^{ll}} \,, \label{eq:delIB_def}
\ee
where $m_l \equiv (m_u + m_d) / 2$.

Table~\ref{tab:amu_IB} summarizes the isospin-breaking shifts on the $N_f=2+1+1$ and $N_f=1+1+1+1$ ensembles, both before and after rescaling the Taylor coefficients.   As expected, we do not observe any significant difference between the two ensembles.   
The leading sea isospin-breaking contributions to \amu\ are quadratic in the difference $(m_d - m_u)$; taking $\Lambda_{\rm QCD} = 300$~MeV gives a rough power-counting estimate of their size as $(m_u -  m_d)^2/\Lambda_{\rm QCD}^2 \sim 0.01\%$.  The differences in \amu\ are smaller with rescaling because the valence-quark-mass dependence is milder. 

The errors on the shifts in Table~\ref{tab:amu_IB} stem primarily from the two-point correlator fits. The parametric errors from the lattice spacing are about a percent before rescaling, and are twenty times smaller with rescaling.  The parametric errors from the current renormalization factor are  $\sim$0.2\%.  The uncertainty due to the use of Pad{\'e} approximants, which we take to be the difference between \amu\ obtained with the $[3,3]$ and $[3,2]$ approximants, is about a percent.  The 2.7\% uncertainty on the ratio $m_u/m_d$ in Ref.~\cite{Basak:2016jnn} stems largely from the estimate of electromagnetic effects, and leads to errors of about 2\% and 1\% on the physical up- and down-quark masses, respectively.  Propagating the tuned quark-mass uncertainties to the physical \amu\ using the measured slope of \amu\ with respect to valence-quark mass changes the shifts in Table~\ref{tab:amu_IB} by $\sim 0.2$--$0.3\times 10^{-10}$ ($\ltapprox 0.1\times 10^{-10}$) without (with) rescaling.  Finally, the leading finite-volume and discretization effects, which arise from one-loop diagrams with $\pi\pi$ intermediate states, cancel in $\Delta a_\mu^{\rm HVP}$ because the charged pions in the loop are sensitive to the average of the up- and down-quark masses.  Higher-order contributions are suppressed by $m_{ud}/\Lambda_\chi \sim 1\%$, where $\Lambda_\chi$ is a typical chiral perturbation theory scale.   We therefore estimate the systematic uncertainties in the shifts in Table~\ref{tab:amu_IB} due to finite-volume and discretization effects to be 1\% times the leading contributions, or $0.5 \times 10^{-10}$.

Because the sea-quark-mass dependence of \amu\ is tiny, we can compare the shift in the ``direct" points in Fig.~\ref{fig:amu_vs_mval} to the valence-quark-mass dependence observed in Ref.~\cite{Chakraborty:2016mwy}, which analyzes several isospin-symmetric MILC HISQ ensembles at three lattice spacings and with a wide range of pion masses.
Figure~3 of that work shows that the ``raw" data for \amu\ are approximately linear in $m_q$ from $M_\pi \sim 300$~MeV down to the physical value with a slope that is independent of the lattice spacing.
We can therefore estimate the change in \amu\ that would result from varying $m_q$ between $m_u$ and $m_d$ from the unphysically heavy data in Ref.~\cite{Chakraborty:2016mwy}, and find a value consistent with the difference obtained from our fully physical calculation here.

The shifts in Table~\ref{tab:amu_IB} only include contributions from quark-connected diagrams, with quark-disconnected contributions  expected to be suppressed by $1/N_c$.   We estimate the quark-disconnected contribution to the strong-isospin-breaking correction from one-loop $\pi\pi$ diagrams, which are especially sensitive to changes in the quark masses, within finite-volume chiral perturbation theory.  Including the effect of taste splittings between the sea pions, which reduce the isopsin-breaking shift, we obtain for the $\pi\pi$-loop contribution $0.7 \times 10^{-10}$.   To account for resonance and higher-order contributions, we take about three times this value, or $3 \times 10^{-10}$, as the uncertainty on the isospin-breaking shifts in Table~\ref{tab:amu_IB} from missing quark-disconnected contributions.    This conservative error estimate is approximately the size of the full quark-disconnected contribution to \amu\ obtained by the BMW Collaboration on their coarsest ensemble with $a\approx 0.13$~fm and similar taste splittings~\cite{Borsanyi:2017zdw}; we expect the quark-disconnected contribution to the strong-isospin splitting to be smaller.

We obtain our final results for the relative correction \DeltaIB\ by averaging the values on the two ensembles.

\begin{table}[tb]
    \caption{Shift in \amu\ from the isospin-symmetric to the physical valence-quark masses calculated on the ensembles in Table~\ref{tab:ensembles}.  Results are shown both without and with rescaling the Taylor coefficients.  As explained in the text, the numbers within  a column should agree, but the two columns can (and should) differ. Errors shown include statistics and all systematic uncertainties. \vspace{1mm}}
    \label{tab:amu_IB}
\begin{ruledtabular}
\begin{tabular}{lc@{\quad}c} 
& \multicolumn{2}{c}{$10^{10}\Delta a_\mu^{{\rm HVP},m_u\neq m_d}$} \\
$N_f$ & direct & with $E_0$ rescaling \\ 
\hline
2+1+1 & +7.7(3.7) & +1.9(4.0) \\
1+1+1+1 & +9.0(2.3) & +2.3(2.5) \\
\end{tabular}
\end{ruledtabular}
\end{table}

\section{Result and outlook}
\label{sec:conclusions}

We obtain for the relative strong isospin-breaking correction to the light-quark connected contribution to the muon $g-2$ hadronic vacuum polarization
\begin{numcases}{\delta a_\mu^{{\rm HVP,} m_u \neq m_d}  = }
+1.5(7)\% & \mbox{direct,} \label{eq:DeltaIB_Raw}\\
+0.4(7)\% & \mbox{with $E_0$ rescaling,\qquad} \label{eq:DeltaIB_RS}
\end{numcases}
where the errors include Monte Carlo statistics and all systematics.   
Our result without rescaling the Taylor coefficients is consistent with phenomenological estimates of the dominant isospin-breaking contribution from $\rho$--$\omega$ mixing using $e^+e^- \to \pi^+\pi^-$ data~\cite{Wolfe:2010gf,Jegerlehner:2011ti,Jegerlehner:2017gek}, $\Delta a_\mu^{\rho-\omega{\rm \ mix.}} \sim 2$--5$\times 10^{-10}$, and chiral perturbation theory~\cite{Cirigliano:2002pv}, $\Delta a_\mu^{\rho-\omega{\rm \ mix.}} \sim 6 \times 10^{-10}$, although $\rho$--$\omega$ mixing will also include effects from quark-line disconnected diagrams that we 
do not consider here.
Recent exploratory lattice-QCD calculations obtain somewhat smaller estimates for the relative strong isospin-breaking correction of roughly 0.2\%--0.6\% for $M_\pi \gtapprox 340$~MeV~\cite{Giusti:2017ier,Boyle:2017gzv}.
We cannot directly compare our result in Eq.~(\ref{eq:DeltaIB_Raw}) with these values, however, because they were obtained with unphysically heavy pions and do not yet include systematic uncertainties.\footnote{Since our paper appeared, the RBC/UKQCD Collaboration obtained a new result for the strong-isospin-breaking shift at the physical pion mass of $10.6(8.0) \times 10^{-10}$~\cite{Blum:2018mom}, which agrees with our ``direct" values in Table~\ref{tab:amu_IB}.}

The percentage shifts in Eqs.~(\ref{eq:DeltaIB_Raw}) and~(\ref{eq:DeltaIB_RS}) can be used to correct any existing or future result for the connected contribution to the hadronic vacuum polarization obtained with degenerate light quarks.  Results for \amu\ obtained without rescaling the Taylor coefficients should be corrected using Eq.~(\ref{eq:DeltaIB_Raw}); this applies to most recent lattice-QCD calculations.  Equation~(\ref{eq:DeltaIB_RS}) should be used to correct \amu\ when $E_0$ rescaling is employed. 

We have performed the first direct calculation of the strong-isosopin-breaking correction to \amu\ at the physical up- and down-quark masses.  We obtain an uncertainty on the relative correction of 0.7, which is smaller, and also more reliable, than the $\sim 1\%$ phenomenological estimate used in recent lattice-QCD calculations with equal up- and down-quark masses~\cite{Chakraborty:2016mwy,Borsanyi:2016lpl,BMW17}.
Thus, it reduces a significant source of uncertainty in \amu, and is a crucial milestone towards a complete {\it ab-initio} lattice-QCD calculation of the hadronic contributions to $a_\mu$ with the sub-percent precision needed by the Muon $g-2$ and planned J-PARC experiments.   

To improve our results in Eqs.~(\ref{eq:DeltaIB_Raw}) and~(\ref{eq:DeltaIB_RS}), we will include quark-disconnected contributions, which are the dominant source of uncertainty, in a future work.   We will also calculate directly the electromagnetic correction to \amu\ using dynamical QCD+QED gauge-field configurations to be generated soon by the MILC Collaboration with quarks, gluons, and photons in the sea~\cite{Zhou:2014gga}.

\begin{acknowledgments}

We thank John Campbell, Vera G{\"u}lpers, Fred Jegerlehner, Laurent Lellouch, and Silvano Simula for useful discussions.

Computations for this work were carried out with resources provided by the USQCD Collaboration, the National
Energy Research Scientific Computing Center and the Argonne Leadership Computing Facility, which are funded
by the Office of Science of the U.S.\ Department of Energy; and with resources provided by the National
Institute for Computational Science and the Texas Advanced Computing Center, which are funded through the
National Science Foundation's Teragrid/XSEDE Program.  Computations were also carried out on the Darwin Supercomputer at the DiRAC facility, which is jointly funded by the U.K.\ Science and Technology Facility Council, the U.K.\ Department for Business, Innovation and Skills, and the Universities of Cambridge and Glasgow.  This work utilized the RMACC Summit supercomputer, which is supported by the National Science Foundation (awards ACI-1532235 and ACI-1532236), the University of Colorado Boulder, and Colorado State University. The Summit supercomputer is a joint effort of the University of Colorado Boulder and Colorado State University.  This research is part of the Blue Waters sustained-petascale computing project, which is supported by the National Science Foundation (awards OCI-0725070 and ACI-1238993) and the state of Illinois. Blue Waters is a joint effort of the University of Illinois at Urbana-Champaign and its National Center for Supercomputing Applications.

This work was supported in part by the U.S.\ Department of Energy under grants
No.~DE-AC05-06OR23177 (B.C.),
No.~DE{-}SC0010120 (S.G.), 
No.~DE{-}SC0015655 (A.X.K.), 
No.~DE{-}SC0009998~(J.L.), 
No.~DE{-}SC0010005 (E.T.N.), 
No.~DE-FG02-13ER41976 (D.T.),
by the U.S.\ National Science Foundation under grants
PHY14-17805~(J.L.), 
PHY14-14614 (C.D., A.V.), 
PHY13-16222 (G.P.L.), 
PHY12-12389~(Y.L.),
and PHY13-16748 and PHY16-20625 (R.S.); 
by the Royal Society, STFC and Wolfson Foundation (C.T.H.D., D.H., J.K.);
by the MINECO (Spain) under grants FPA2013-47836-C-1-P and FPA2016-78220-C3-3-P (E.G.); 
by the Junta de Andaluc{\'i}a (Spain) under grant No.~FQM-101 (E.G.)
by the Fermilab Distinguished Scholars Program (A.X.K.);
by the German Excellence Initiative and the European Union Seventh Framework Program under grant agreement No.~291763 as well as the European Union's Marie Curie COFUND program (A.S.K.);
by the Blue Waters PAID program (Y.L.);
and by the U.K.\ STFC under grants  ST/N005872/1 and ST/P00055X/1 (C.M.).

Brookhaven National Laboratory is supported by the U.S. Department of
Energy under contract DE-SC0012704.
Fermilab is operated by Fermi Research Alliance, LLC, under Contract No.\ DE-AC02-07CH11359 with the United States Department of
Energy, Office of Science, Office of High Energy Physics.
The United States Government retains and the publisher, by accepting the article for publication, acknowledges that the United
States Government retains a non-exclusive, paid-up, irrevocable, world-wide license to publish or reproduce the published form of this manuscript, or allow others to do so, for United States Government purposes.

\end{acknowledgments} 

\bibliographystyle{apsrev4-1} %%% physical review (up to date)
\bibliography{./bibliography}

\end{document}